\documentclass[12pt]{iopart} 

\usepackage{amsfonts}
\usepackage{amssymb}
\usepackage{graphicx}
\usepackage{subfig} 
\usepackage{bm}
\usepackage{latexsym}

\def\a{\alpha}
\def\l{\ell}
\def\e{\epsilon}
\def\bl{{\bar\ell}}
\def\av#1{\left\langle#1\right\rangle}

\def\comment#1{}
\newcommand{\be}{\begin{equation}}
\newcommand{\ee}{\end{equation}}

\begin{document}

\title{Deterministic reaction models with power-law forces}
\author{Daniel ben-Avraham$^{1}$, Oleksandr Gromenko$^1$, and Paolo Politi$^{2,3}$}
\address{$^1$ Physics Department, Clarkson University, Potsdam, NY 13699-5820, USA}
\address{$^2$ Istituto dei Sistemi Complessi, Consiglio Nazionale
delle Ricerche, Via Madonna del Piano 10, 50019 Sesto Fiorentino, Italy}
\address{$^3$ INFN Sezione di Firenze, via G. Sansone 1, 50019 Sesto Fiorentino, Italy}
\eads{
\mailto{benavraham@clarkson.edu},
\mailto{gromenko@clarkson.edu},
\mailto{paolo.politi@isc.cnr.it}
} 
\begin{abstract}
We study a one-dimensional particles system, in the overdamped limit,
where nearest particles attract with a force inversely proportional to a power $\a$ of their distance 
and coalesce upon encounter.  The detailed shape of the distribution function for the gap between
neighbouring particles serves to discriminate between different laws of attraction.
We develop an exact Fokker-Planck approach for the infinite hierarchy of distribution functions
for multiple adjacent gaps and solve it exactly, at the mean-field level,
where correlations are ignored.  The crucial role of correlations and their effect on the gap distribution
function is explored
both numerically and analytically.  Finally, we analyse  a random input of particles, which
results in a stationary state where the effect of correlations is largely diminished.
\end{abstract}


\maketitle  

\section{Introduction}

In a recent paper~\cite{PhysD}, we have mapped the late-time evolution of the conserved Kuramoto-Sivashinsky (CKS) equation~\cite{CKS},
\be
u_t = - (u + u_{xx} + u_x^2)_{xx}\,, 
\ee
used to describe, e.g., the dynamics of meandering steps in crystal growth
or that of wind driven sand dunes, to a reacting particles system.
The long-time profile $u(x,t)$ is made up of pieces of a universal parabola
(solution of the equation $u + u_{xx} + u_x^2=c$),
 joining in regions
of vanishing size and diverging curvature whose locations define the profile without ambiguity,
and that one may associate with `particles'.   
The projected particles on the $x$-axis 
form an overdamped one-dimensional system; nearest particles attract one another with a
force inversely proportional to their distance, and particles coalesce irreversibly upon encounter.
This process gives rise to a reduction in the particle density, i.e., to a coarsening process.

In this paper, we generalize the above reaction model to the case where the force between 
nearest particles, separated by a distance $\l$, is proportional to $1/\l^{\a}$ ($\a=1$ corresponding
to the CKS equation).  Reacting particles systems such as this, evolving deterministically, occur often enough but have been studied sparingly.  Coarsening 
 in spin systems has been studied for exponential forces~\cite{kawasaki} between the kinks and for
power-law forces (between {\em all} particle pairs)~\cite{krapivsky,bray}.  The deterministic Kardar-Parisi-Zhang equation corresponds to an interacting particles system~\cite{kpz}, and the (non conserved) KS equation
was shown to be equivalent to coalescing particles traveling ballistically~\cite{krug}, as is also the case
for the Burgers equation at high Reynolds numbers (shocks representing particles)~\cite{burgers}. Similar
ballistic systems have been studied by Redner et al.~\cite{redner}.  Simple deterministic reaction schemes, 
such as the ``cut-in-half" model, have been devised by Derrida et al.~\cite{derrida}, but exact results are few
even in those ideal cases. 

The aim of this paper is to give a thorough description of these models with varying $\alpha$.
The time-dependence of coarsening can be easily found using dimensional analysis, however, it 
provides too weak a characterization of the kinetics (e.g., diffusion-limited one-species coalescence coarsens in the same way as our model for $\a=1$).  We therefore 
focus on the distribution function $g(\l,t)$ for the gap $\l$ between nearest particles.  We derive an exact hierarchy of Fokker-Planck equations for the
probability density functions (pdf) $\{\rho_m\}_{m=1}^\infty$, for finding $m$ consecutive gaps of given sizes at time $t$.
Truncating this hierarchy at the mean-field level, ignoring correlations between adjacent gaps, we solve for the
(approximate) single gap distribution function.  The shape of this pdf is a much better discriminant between various
models, and we focus in particular on the exponent $\beta$ characterizing the small gap behavior, $g(\l)\sim\l^\beta$.  The role of correlations is explored numerically and analytically, using the Fokker-Planck equations to derive exact constraints that the gap distributions must satisfy
when correlations are fully taken into account.  Finally, we test the mean-field approximation for the case where
a homogeneous random input of particles produces a steady state that is considerably less affected by correlations.

Apart from the intrinsic interest in deterministic reaction models, our particular system might also be relevant to the dynamics of vicinal surfaces
undergoing step-bunching instability~\cite{bunching} --- a process where parallel straight steps
do not move uniformly but rather bunch together in groups separated by
large terraces. In this picture, particles represent steps. One of the several ways that
steps interact is through diffusion, giving rise to an effective attraction
 decaying
as $1/(b + \l)$, where $b$ is some characteristic length. For large terraces, the
model $\a=1$ is recovered, while the opposite limit, $\l\ll b$, corresponds to our model with
$\a=-1$.

\section{The Model}

Consider an infinite system of particles on the line, located at $\{x_i(t)\}_{i=-\infty}^{\infty}$.
The system is overdamped, and nearest particles attract one another with a force inversely proportional
to a power $\a$ of the distance:
\begin{equation}
\label{dx/dt}
\frac{dx_i}{dt}=\frac{A}{\a}\left(\frac{1}{(x_{i+1}-x_i)^\a}-\frac{1}{(x_i-x_{i-1})^\a}\right)\,.
\end{equation}
The prefactor $A$ carries units of $({\rm length})^{\a+1}/({\rm time})$ (or ${\rm L}^{\a+1}/{\rm T}$, symbolically), and $1/\a$ has been introduced to enable
discussion of negative $\a$, as well as the limit $\a\to0$. 

From (\ref{dx/dt}) one finds that the gaps between particles, $\l_i=x_{i+1}-x_{i}$, obey
\begin{equation}
\label{dl/dt}
\frac{d\l_i}{dt}=\frac{A}{\a}\left(\frac{1}{\l_{i-1}^\a}-\frac{2}{\l_i^\a}+\frac{1}{\l_{i+1}^\a}\right)\,.
\end{equation}
The particles coalesce upon encounter, or analogously a gap is removed from the system once it has shrunk
to zero.  Thus, the number of particles decreases and  the typical gap between 
particles grows as the system coarsens with time. 

\subsection{Linear Stability Analysis}
To analyse the system's stability let us perturb a uniform configuration,
\begin{equation}
\l_i = \av{\l} + \e_i\,,\qquad
\l_i^{-\alpha} \simeq  \av{\l}^{-\alpha} \left( 1-\alpha {\e_i\over\av{\l}}\right)\,,
\end{equation}
and substitute in (\ref{dl/dt}) to yield
\begin{equation}
\dot\e_i = -{A\over\av{\l}^{\alpha+1}} (\e_{i+1}+\e_{i-1}-2\e_i)\,,
\end{equation}
or, upon passing to the continuum limit (for large $i$ and $t$), $\e_i\to\e(x,t)$,
\begin{equation}
\frac{\partial}{\partial t}\e(x,t)= - {A\over\av{\l}^{\alpha+1}} \frac{\partial^2}{\partial x^2}\e(x,t)\,.
\end{equation}
Thus the perturbations obey a diffusion equation with {\em negative} coefficient, and the system is unstable regardless of the sign of $\a$: the prefactor $1/\a$ introduced in (\ref{dx/dt}) ensures this outcome.  For $\a>0$
the attractive forces between particles separated by short gaps grow ever stronger, resulting in  the eventual coalescence of the particles, and corresponding to  singularity buildups in the diffusion equation.  For $\a<0$, the forces are {\em repulsive} and short gaps shrink into singularities due to the dominating forces of the larger surrounding gaps.  In the demarcating limit of $\a\to0$ the forces depend logarithmically upon the distance:
\begin{equation}
\frac{dx_i}{dt}=A\ln\frac{x_i-x_{i-1}}{x_{i+1}-x_i}\,;\qquad \frac{d\l_i}{dt}=A\ln\frac{\l_i^2}{\l_{i-1}\l_{i+1}}\,.
\end{equation}

\subsection{Simulations}
Simulations have been performed by numerical integration of Eq.~(\ref{dx/dt})
for systems of $10^4$ particles (for $\a>0$) and $10^5$ particles ($\a<0$).  
The calculations were carried out on a Linux AMD64 cluster comprising 32 processors.  
Data were obtained by averaging over
thousands of independent runs. The long-time asymptotic limit is reached sooner for
positive $\a$, so typically the runs proceeded until the number of particles
dwindled to $500$--$700$ for $\a>0$, or $50$--$90$ for $\a<0$.

\section{Dimensional analysis}
Numerical integration of the equations of motion, starting from different initial conditions, suggests that the original state
of the system has no effect on its long-time asymptotic behavior.  In the long-time regime, therefore, the only 
dimensional physical parameters determining the typical gap $\bl$ (a length, L) are the elapsed time $t$ (T)
and the constant $A$ (${\rm L}^{\a+1}/{\rm T}$), so one expects
\begin{equation}
\bl\sim(At)^{1/(1+\a)}\,.
\label{c_law}
\end{equation}
This is indeed confirmed by numerical simulations.

This scaling argument breaks down as $\a$ becomes smaller than $-1$.  As $\a\to-1$ coarsening proceeds ever faster and takes place at a {\em finite} time (even in an infinite system) in the limit $\a=-1$,
and the initial configuration becomes then important.  Independence from initial conditions, however, is a prerequisite for the scaling behavior that we assume (and find numerically) in order to analyse the gap distribution in Section~\ref{gap}.  Accordingly, we limit the discussion
to $\a>-1$.

If in addition to the coalescence reaction particles are input randomly, at rate $R$ per unit length per unit time (or $1/{\rm LT}$),
the system achieves a steady state that is independent of its initial condition.  The typical distance at the steady state is then dependent upon $A$ and $R$ alone, so
\begin{equation}
\bl_s\sim\left(\frac{A}{R}\right)^{1/(2+\a)}\,.
\end{equation}
Physically, the characteristic time for the shrinking of $\bl_s$, $\bl_s^{1+\a}/A$, is then balanced by
the time $1/\bl_s R$ for getting a new particle into the interval.
For the important case of $\a=1$, the exponents $\bl\sim t^{1/2}$ and $\bl_s\sim R^{-1/3}$ agree exactly with the well known cases of diffusion-limited coalescence and diffusion-limited coalescence with input (where the particles diffuse freely without damping and without an attractive force)~\cite{coal}.

\section{Gap distributions and Fokker-Planck approach}
\label{gap}

\begin{figure}[ht]
\vspace*{0.cm}
\includegraphics*[width=0.9\textwidth]{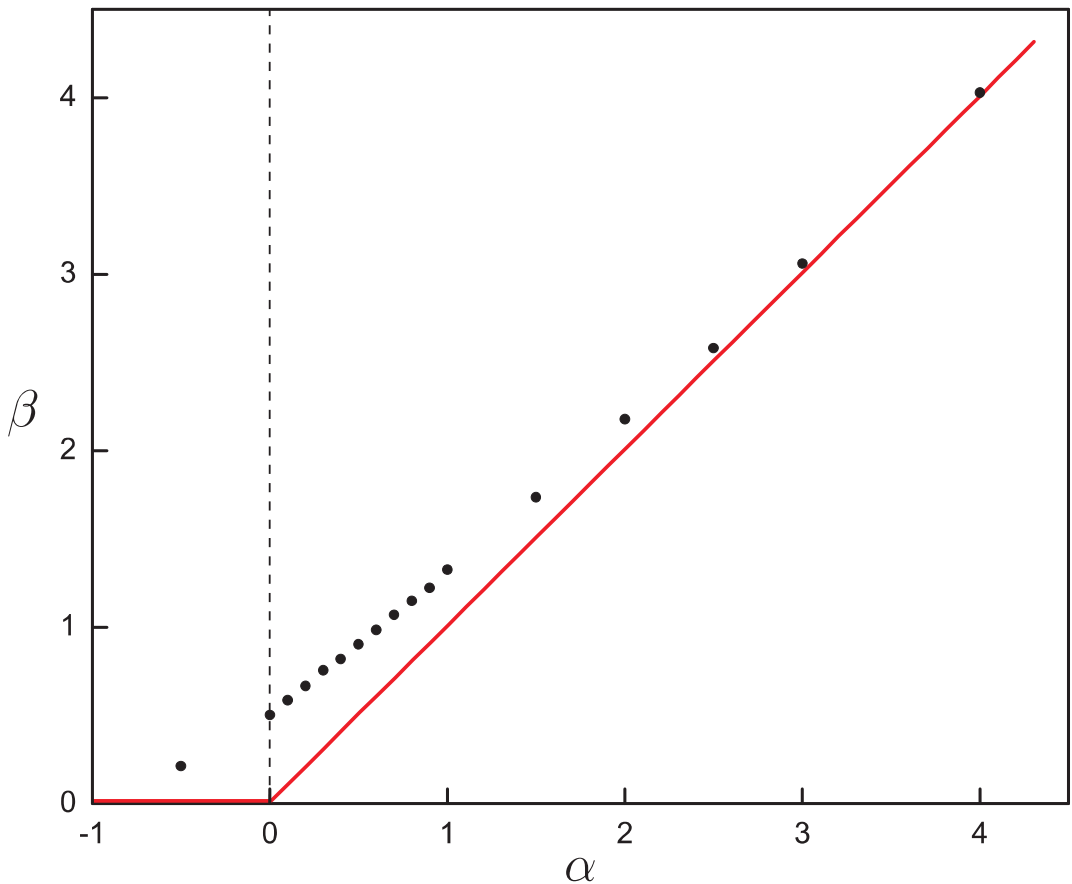}
\caption{The exponent $\beta$, that describes the small gap behavior of the gap distribution function, $\rho(\l,t)\sim \l^{\beta}$, as a function of the force exponent $\a$ in our deterministic reaction model.  The solid line indicates the mean-field prediction of Eq.~(\ref{small_z}), $\beta=\alpha$.}
\label{fig.beta}
\end{figure}

We see that the coarsening kinetics is not sufficient to distinguish between such different systems as diffusion-limited coalescence, annihilation, and our model with $\a=1$.  A full characterization of a particles system on the line is provided by the hierarchy of joint pdfs for the multiple-gap correlation functions $\rho_m(\l_1,\l_2,\dots,\l_m,t)$
--- the probability of finding $m$ consecutive gaps of lengths $\l_1,\l_2,\dots,\l_m$ at time $t$. This is fully
equivalent to the more traditional hierarchy of multiple-point correlation functions $p_k(x_1,x_2,\dots,x_k,t)$
for finding $k$ particles at $x_1,x_2,\dots,x_k$ at time $t$~\cite{brunet}.  In the following, we develop a Fokker-Planck approach for obtaining the $\rho_m$.  For practical purposes, however, the single gap distribution $\rho_1\equiv\rho(\l,t)$ provides already with enough detail to distinguish between various reaction models.  For example, the tail of the distribution falls off as a gaussian, $\rho\sim\exp(-c_1\l^2/t)$, for diffusion-limited coalescence, but as
a stretched exponential, $\exp(-c_2\l^{3/2})$, for coalescence with input, while for small $\l$ one finds $\rho\sim\l^{\beta}$ with 
$\beta=1$ in both these cases
but $\beta\approx1.3$ for our model with $\a=1$.  For other values of $\a$ one finds that $\beta$ has a non trivial dependence upon $\a$ (Fig.~\ref{fig.beta}).

Let $n_m(\l_1,\l_2,\dots,\l_m,t)d\l_1d\l_2\cdots d\l_m$ denote the number of $m$ consecutive gaps of lengths between
$\l_j$ and $\l_j+d\l_j$ ($j=1,2,\dots,m$), at time $t$.  $m=1$ corresponds to the single
gap distribution $n_1 d\l \equiv n(\l,t)d\l$ --- the number of gaps between length $\l$ and $\l+d\l$, at time $t$.  The total number of gaps (or particles),
\begin{equation}
N(t)=\int_0^{\infty}n(\l,t)\,d\l\sim1/\bl\sim t^{-1/(1+\a)}\,,
\end{equation}
decreases with time,
since the total length of the system $\Lambda\sim N(t)\bl$ is constant.
Likewise, the normalization for $m>1$  
is
\begin{equation}
\int_0^{\infty}\cdots\int_0^{\infty}n_m(\l_1,\dots,\l_m,t)\,d\l_1\cdots d\l_m=N(t)\,,
\end{equation}
assuming periodic boundary conditions. For free boundary conditions the result is $N(t)-m+1$, but $(m-1)$ can be neglected in the thermodynamic limit of $N(0)\to\infty$.
The joint probability density, $\rho_m$, is related to $n_m$ via the $N(t)$ normalization:
\begin{equation}
\rho_m(\l_1,\dots,\l_m,t)=N(t)^{-1}n_m(\l_1,\dots,\l_m,t)\,.
\end{equation}

$n_m$~experiences a systematic drift, due to the probability ``current" ${\bf J}$; 
$J_j=n_mv(\l_{j-1},\l_j,\l_{j+1})$, arising from the deterministic evolution
of each of the gaps $\l_j$ at the rate of $d\l_j/dt= v(\l_{j-1},\l_j,\l_{j+1})= 
(1/2\a)(1/\l_{j-1}^\a-2/\l_j^\a+1/\l_{j+1}^\a)$, as given by
Eq.~(\ref{dl/dt}) and absorbing a factor of $2A$ in $t$.  Then, using the notation $\vec\l_m\equiv (\l_1,\dots,\l_m)$, we can write the Fokker-Planck
equation
\begin{eqnarray}
\label{nm}
\fl \frac{\partial}{\partial t}n_m(\vec\l_m,t)=-\sum_{j=2}^{m-1}\frac{\partial}{\partial\l_j}
\left[ 
v(\l_{j-1},\l_j,\l_{j+1})
n_m(\vec\l_m,t)\right]\nonumber\\
\lo-\frac{\partial}{\partial\l_1}
\int_0^{\infty}
v(\l',\l_1,\l_2)
n_{m+1}(\l',\vec\l_m,t)\,d\l'\nonumber\\
\lo-\frac{\partial}{\partial\l_m}
\int_0^{\infty}
v(\l_{m-1},\l_m,\l'')
n_{m+1}(\vec\l_m,\l'',t)\,d\l''\nonumber\\
\lo-\sum_{j=1}^{m-1}\lim_{\l_j'\to0}\left[
v(\l_j,\l'_j,\l_{j+1})
n_{m+1}(\l_1,\dots,\l_j,\l_j',\l_{j+1},\dots,\l_m,t)\right]\,.
\end{eqnarray}
Note that since the expression for $d\l/dt$ requires knowledge of the gaps lengths to the left and right of $\l$, the
drift for $\l_1$ (the second term on the rhs) forces us to consider ${m+1}$ gaps, including $\l'$ (the neighbour
of $\l_1$ to its left), and similarly for $\l_m$ (the third term on the rhs) which requires the gap $\l''$ to its right.  The last term denotes the creation of $n_m$
when the gap $\l_j'$ shrinks and disappears (due to coalescence) from the $(m+1)$-gap configuration
$\l_1,\l_2,\dots,\l_j,\l_j',\l_{j+1},\dots,\l_m$.  Thus the Fokker-Planck equation for $n_m$ requires $n_{m+1}$, resulting in an infinite hierarchy of equations.

\subsection{Scaling}
\begin{figure}[ht]
\vspace*{0.cm}
\includegraphics*[width=0.9\textwidth]{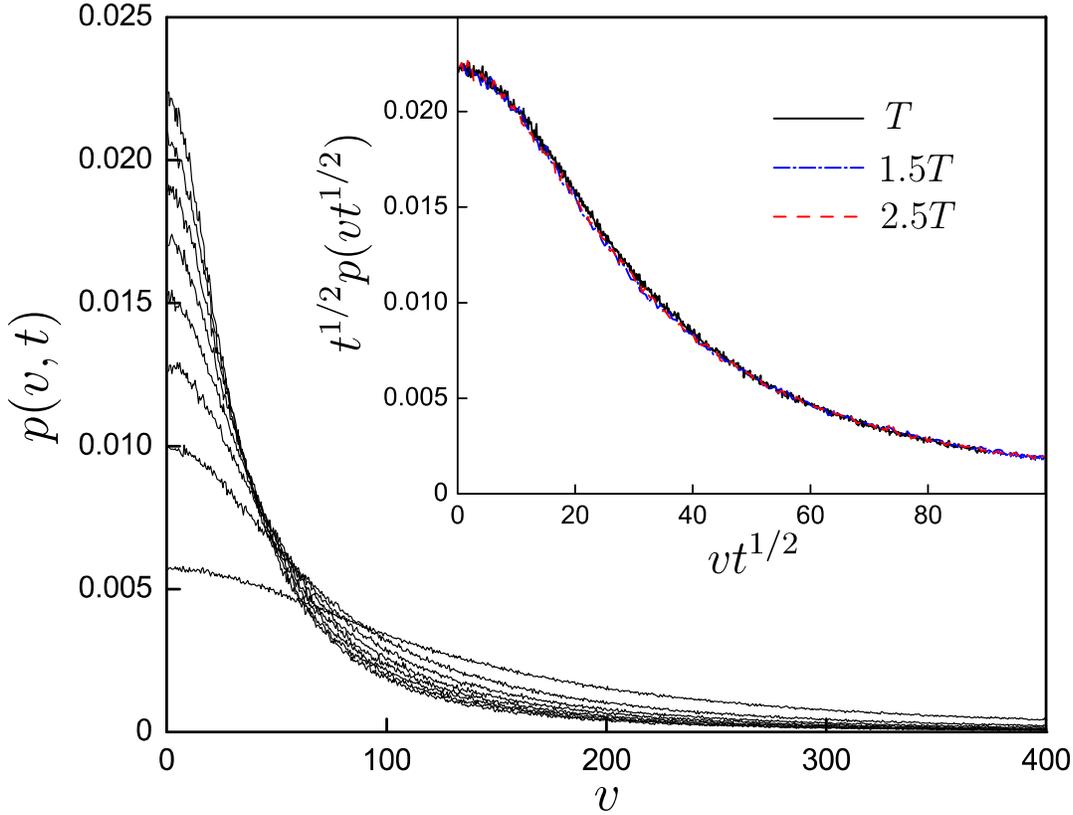}
\caption{Speed distribution function, shown for several times, for the case of $\a=1$.  Inset: Collapse of the pdfs is shown for three
different times after the scaling regime set in (by time $T$ the initial number of particles has been reduced by a factor of 37), using the relation $\bl\sim t^{1/2}$, valid for $\a=1$.}
\label{speed.fig}
\end{figure}

A useful simplification that gets rid of the time variable $t$ results from the scaling assumption
\begin{equation}
\label{scaling}
\fl n_m(\l_1,\dots,\l_m,t)=N(t)\rho_m(\l_1,\dots,\l_m,t)=N(t)\bl^{-m}h_m(z_1,\dots,z_m);\quad z_j
\equiv \ell_j/\bl\,.
\end{equation}
In~\cite{PhysD} we have presented simulation results supporting this scaling form for $m=1$, or the single-gap pdf.
For $m=2$ we can test the scaling hypothesis in a roundabout way, by focusing on the distribution function
of the particles' speeds
\be
p(v,t)=\int_0^{\infty}\int_0^{\infty}\delta\left(v-|\case{1}{\a}(\l'^{-\a}-\l''^{-\a})|\right)\rho_2(\l',\l'',t)\,d\l'd\l''\,.
\ee
A simple change of variables now shows that (\ref{scaling}) implies
the scaling of the speed pdf:
 $p(v,t)=\bl^{\a}p(v\bl^{\a})$.  In Fig.~\ref{speed.fig} we confirm that this scaling indeed
holds, for the case of $\a=1$, giving us some confidence in the postulated scaling form (\ref{scaling}) for general $m$.

Using  (\ref{scaling}) and the fact that $N(t)\sim1/\bl$, the Fokker-Planck equation reads,
\begin{eqnarray}
\label{hm}
\fl \a(m+1)h_m(z_1,\dots,z_m)+\a\sum_{j=1}^mz_j\frac{\partial}{\partial z_j}h_m(z_1,\dots,z_m)\\
\lo=\sum_{j=2}^{m-1}\frac{\partial}{\partial z_j}
\left[\left(\frac{1/2}{z_{j-1}^\a}-\frac{1}{z_j^\a}+\frac{1/2}{z_{j+1}^\a}\right)h_m(z_1,\dots,z_m)\right]\nonumber\\
+\frac{\partial}{\partial z_1}
\int_0^{\infty}\left(\frac{1/2}{z'^\a}-\frac{1}{z_1^\a}+\frac{1/2}{z_{2}^\a}\right)h_{m+1}(z',z_1,z_2\dots,z_m)\,dz'\nonumber\\
+\frac{\partial}{\partial z_m}
\int_0^{\infty}\left(\frac{1/2}{z_{m-1}^\a}-\frac{1}{z_{m}^\a}+\frac{1/2}{z''^\a}\right)h_{m+1}(z_1,\dots,z_{m-1},z_m,z'')\,dz''\nonumber\\
+\sum_{j=1}^{m-1}\lim_{z_j'\to0}\left[\left(\frac{1/2}{z_j^\a}-\frac{1}{z_j'^\a}+\frac{1/2}{z_{j+1}^\a}\right)
h_{m+1}(z_1,\dots,z_j,z_j',z_{j+1},\dots,z_m)\right].\nonumber
\end{eqnarray}
In fact, to arrive at this equation it is necessary to assume that ${\dot\bl}\bl^\a$
is a constant (that can be absorbed in the time units, so we take it to be 1).  
This condition is in agreement with the coarsening law found from dimensional analysis,
Eq.~(\ref{c_law}).

\subsection{Truncation scheme and Mean Field solution}
Eq.~(\ref{hm}) is  valid for $m\ge 2$.  This infinite hierarchy of equations can be truncated at
any level $m$ by the usual Kirkwood approximation scheme:
\begin{equation}
h_{m+1}(z_1,\dots,z_{m+1})=\frac{h_m(z_1,\dots,z_m)h_m(z_2,\dots,z_{m+1})}{h_{m-1}(z_2,\dots,z_m)}\,.
\end{equation}
For example, at the lowest level of $m=2$ we can truncate the hierarchy (\ref{hm})  using
\begin{equation}
\label{K2}
h_3(z_1,z_2,z_3,)\approx\frac{h_2(z_1,z_2)h_2(z_2,z_3)}{h(z_2)}\,.
\end{equation}
This yields
\begin{eqnarray}
\label{h2}
\fl 3\a h_2(z_1,z_2)+\a z_1\frac{\partial}{\partial z_1}h_2(z_1,z_2)+\a z_2\frac{\partial}{\partial z_2}h_2(z_1,z_2)\nonumber\\
\lo=\frac{\partial}{\partial z_1}\left\{h_2(z_1,z_2)\left[\frac{1/2}{h(z_1)}\int_0^{\infty}\frac{dz'}{z'^\a}h_2(z_1,z')
  +\left(\frac{1/2}{z_2^\a}-\frac{1}{z_1^\a}\right)\right]\right\}\nonumber\\
+\frac{\partial}{\partial z_2}\left\{h_2(z_1,z_2)\left[\frac{1/2}{h(z_2)}\int_0^{\infty}\frac{dz'}{z'^\a}h_2(z_2,z')
  +\left(\frac{1/2}{z_1^\a}-\frac{1}{z_2^\a}\right)\right]\right\}\nonumber\\
+\lim_{z'\to0}\frac{h_2(z_1,z')h_2(z_2,z')}{h(z')}\left(\frac{1/2}{z_1^\a}-\frac{1}{z'^\a}+\frac{1/2}{z_2^\a}\right)\,,
\end{eqnarray}
which is a closed equation for $h_2$, since $h(z)\equiv h_1(z)=\int_0^{\infty}h_2(z,z')\,dz'$.
This equation is hard to solve, or even to work out its limiting behavior.
%
%
We simplify further by making the {\em mean-field} assumption that adjacent gaps are not correlated, so that
\begin{equation}
h_2(z_1,z_2)=h(z_1)h(z_2)\,.
\end{equation}
Inserting this in (\ref{h2}) and integrating over $z_2$, we obtain 
\begin{equation}
\label{h}
2\a h(z)+\a zh'(z)=\left[\left(\av{\frac{1}{z^\a}}-\frac{1}{z^\a}\right)h(z)\right]'\,,
\end{equation}
where the prime denotes differentiation with respect to $z$.
Thus,
\begin{equation}
\label{hratio}
\frac{h(z)'}{h(z)}=\frac{-2\a+\a z^{-\a-1}}{\a z+z^{-\a}-\av{z^{-\a}}}\,.
\label{eq_MF}
\end{equation}

A naive inspection of Eq.~(\ref{eq_MF}) in the limit of large $z$ leads to the erroneous
conclusion that $h(z) \sim 1/z^2$. The problem is with $\av{z^{-\a}}$ (which must be determined
self-consistently): depending on its value, the denominator on the rhs might vanish, yielding
a singularity that must be taken into account.  Proceeding more carefully, 
we multiply Eq.~(\ref{h}) by $z^k$ and integrate
over $z$, from $0$ to $\infty$, to yield
\begin{eqnarray}
\fl 2\a\av{z^k}+\a z^{k+1}h(z)|_{z=0}^\infty-\a(k+1)\av{z^k}\nonumber\\
\lo=[z^k(\av{z^{-\a}}-z^{-\a})h(z)]_{z=0}^\infty-k\av{z^{-\a}}\av{z^{k-1}}+k\av{z^{k-1-\a}}\,.\nonumber
\end{eqnarray}
Putting $k=0$ we obtain the small-$z$ behavior of $h(z)$:
\begin{equation}
\label{small_z}
h(z)\longrightarrow\cases{
\a z^\a &for $\a>0$,\\
|\a|/\av{z^{|\a|}} &for $\a<0$.}
\label{small_z_MF}
\end{equation}
For $k\to\infty$, and assuming that $h(z)$ has finite support so that the various moments exist and $\lim_{k\to\infty}\av{z^{k+c}}/\av{z^k}=1$, we get, for the unknown ($-\a$)-moment,
\begin{equation}
\label{za}
\av{z^{-\a}}=1+\a\,.
\end{equation}
Armed with this information we now examine the denominator on the rhs of (\ref{hratio}): For $\a>0$, $\a z+z^{-\a}$
achieves a minimum of $1+\a$, at $z=z_c=1$, so that 
the denominator is always positive except at $z=z_c$, where it vanishes.
Meanwhile, the numerator, $-2\a+\a z^{-\a-1}$ is positive at $z<z_m=(1/2)^{1/(1+\a)}$,
and negative for $z>z_m$, (note $z_m<z_c$).  Therefore, $h(z)$ rises as $\a z^{\a}$ (for $z$ near $0$), reaching a maximum at $z=z_m$,
then drops back to zero as $z$ approaches $z_c$. The finite support assumed {\em a priori} is caused by the singularity at $z_c$ and is now seen to be justified.  For $\a<0$ (but $\a>-1$),  $\a z+z^{-\a}$
achieves a maximum of $1+\a$, at $z=z_c=1$, so that 
the denominator is always negative except at $z=z_c$, where it vanishes.
The numerator is negative for $z<z_m$ and positive for $z>z_m$, so
$h(z)$ starts at $|\a|/(1-|\a|)$ at $z=0$, rising to a maximum at $z=z_m$ and drops back to zero towards $z=z_c=1$.

For the case of $\a<0$ one can derive from (\ref{hratio}) the correction to $h(0)$:
\begin{equation}
h(z)\approx h(0)\left(1+\frac{z^{|\a|}}{\av{z^{|\a|}}}\right)=\frac{|\a|}{1-|\a|}\left(1+\frac{1}{1-|\a|}z^{|\a|}\right)\,,
\qquad \a<0\,,
\end{equation}
where we have used (\ref{small_z}) and (\ref{za}) for the last equality.  Numerically, it is impossible to separate between the finite $h(0)$ and the non-analytic correction, as both result in an infinite slope as $z\to0$.  Indeed, the exponent $\beta$ measured for $\a=-1/2$ in Fig.~\ref{fig.beta} was obtained assuming 
$h(0)=0$, yielding only an {\em effective} value.

\begin{figure}[ht]
\vspace*{0.cm}
\includegraphics*[width=0.98\textwidth]{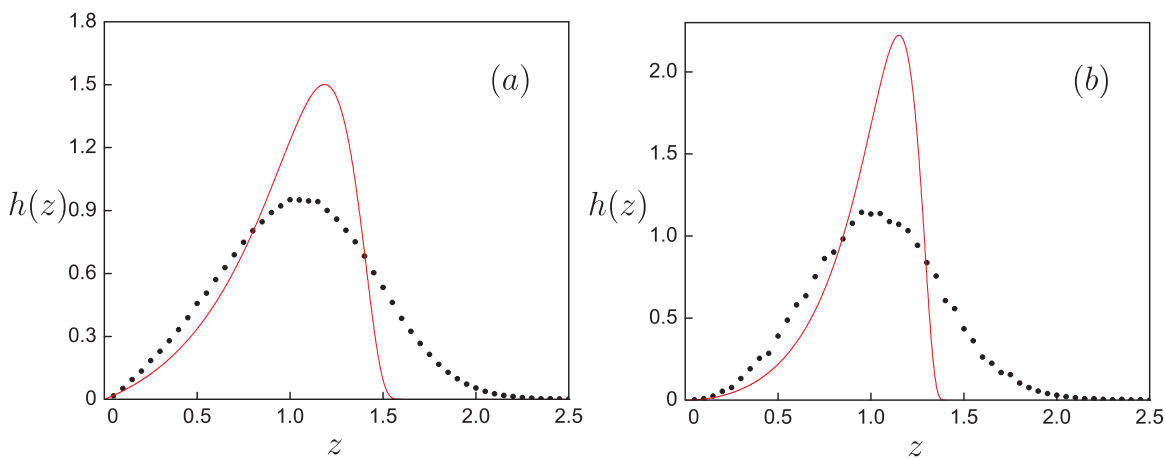}
\caption{Comparison of $h(z)$ from Eqs.~(\ref{hMF1}), (\ref{hMF2}) (solid curves) to actual simulations (symbols), for the cases  of $\a=1$ (a) and $\a=2$ (b). In both cases $z$ has been rescaled so
that $\av{z}=1$.}
\label{halpha}
\end{figure}

\begin{figure}[ht]
\vspace*{0.cm}
\includegraphics*[width=0.9\textwidth]{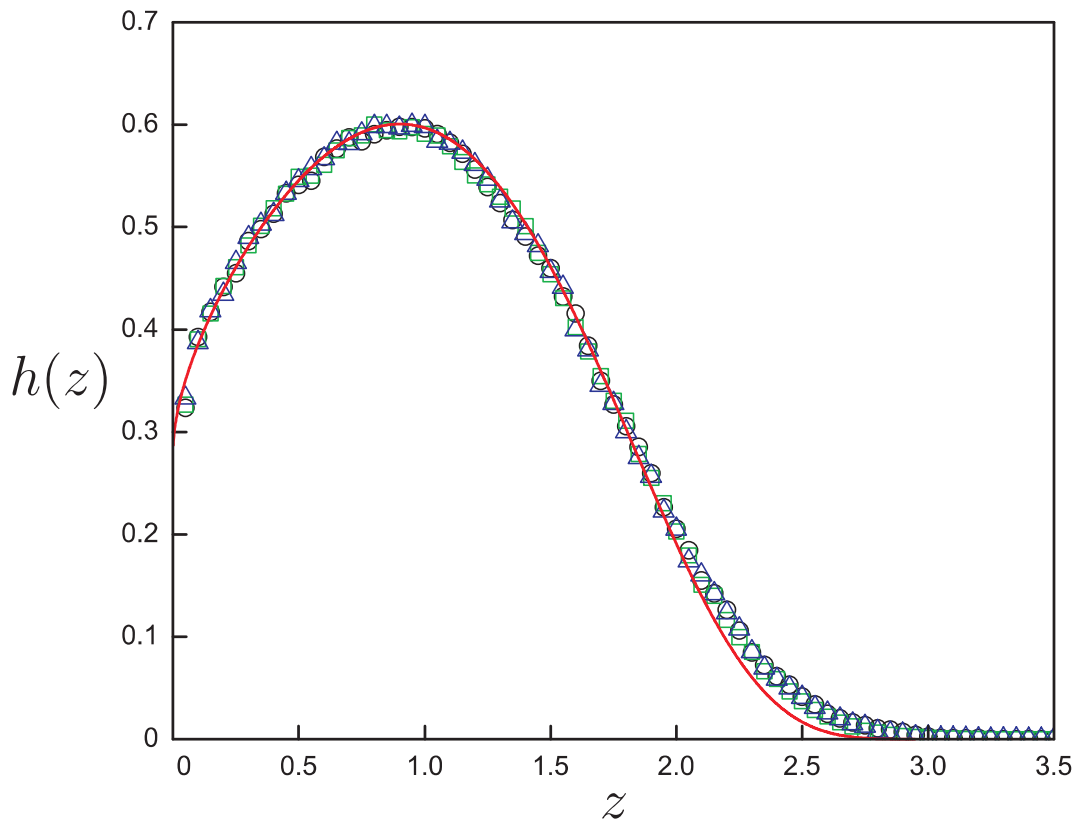}
\caption{Gap distribution for $\a=-1/2$. Simulations  are shown for times $T$ (triangles), $1.5T$ (squares), and $2.5T$ (circles) after the scaling regime had been reached: by time $T$ the particles had been decimated 667-fold. The solid curve represents the mean-field result from Eq.~(\ref{hMF-1/2}).  For the sake of comparison, $z$ has been rescaled to yield $\av{z}=1$.}
\label{halphaneg}
\end{figure}

Finally, it is possible to obtain explicit solutions for several values of $\a$. For example, for the special cases of $\a=1,2$, and $-1/2$ we get
\begin{eqnarray}
\label{hMF1}
\fl h(z)=\frac{z}{(1-z)^3}\,e^{-z/(1-z)}\,,\qquad &0\leq z\leq 1\,;\qquad \alpha=1\,,\\
\label{hMF2}
\fl h(z)=\frac{2z^2}{(1-z)^{26/9}(1+2z)^{10/9}}\,e^{-\frac{2}{3}z/(1-z)}\,,\qquad &0\leq z\leq 1\,;\qquad \alpha=2\,,\\
\label{hMF-1/2}
\fl h(z)=\frac{1}{(1-\sqrt{z})^4}\,e^{-2\sqrt{z}/(1-\sqrt{z})}\,,\qquad &0\leq z\leq 1\,;\qquad \alpha=-1/2\,.
\end{eqnarray}
In all these cases (and others) it is easy to confirm the predicted small $z$ behaviour, the location of the maximum at
$z_m=(1/2)^{1/(1+\a)}$, the finite support to $z<z_c=1$, and that indeed $\av{z^{-\a}}=1+\a$.
In Figs.~\ref{halpha} and \ref{halphaneg} we compare these mean-field results to numerical simulations.
Fig.~\ref{halphaneg} additionally demonstrates that the scaling regime is achieved, numerically, also for 
 $\a<0$, though  this takes a lot longer than for $\a>0$. Reaching scaling from an initial random configuration proved impractical and we had to start the system with all the gaps nearly equal but for  tiny perturbations.

For $\a\gg1$ the forces on particles across narrow gaps is immense and unaffected by neighbouring
particles, therefore one expects that the mean-field prediction for small gaps improves as $\a$ increases (notice indeed the increasing agreement in $\beta$ as $\a\to\infty$, in Fig.~\ref{fig.beta}).  For $\a<0$, on the other hand,
it is the force across unusually large gaps that dominates and is unaffected by neighbouring particles, so there
we expect that mean-field predicts the tail of the distribution increasingly better as $\a$ approaches $-1$. 

\section{The Role of Correlations}
\begin{figure}[ht]
\vspace*{0.cm}
\includegraphics*[width=0.9\textwidth]{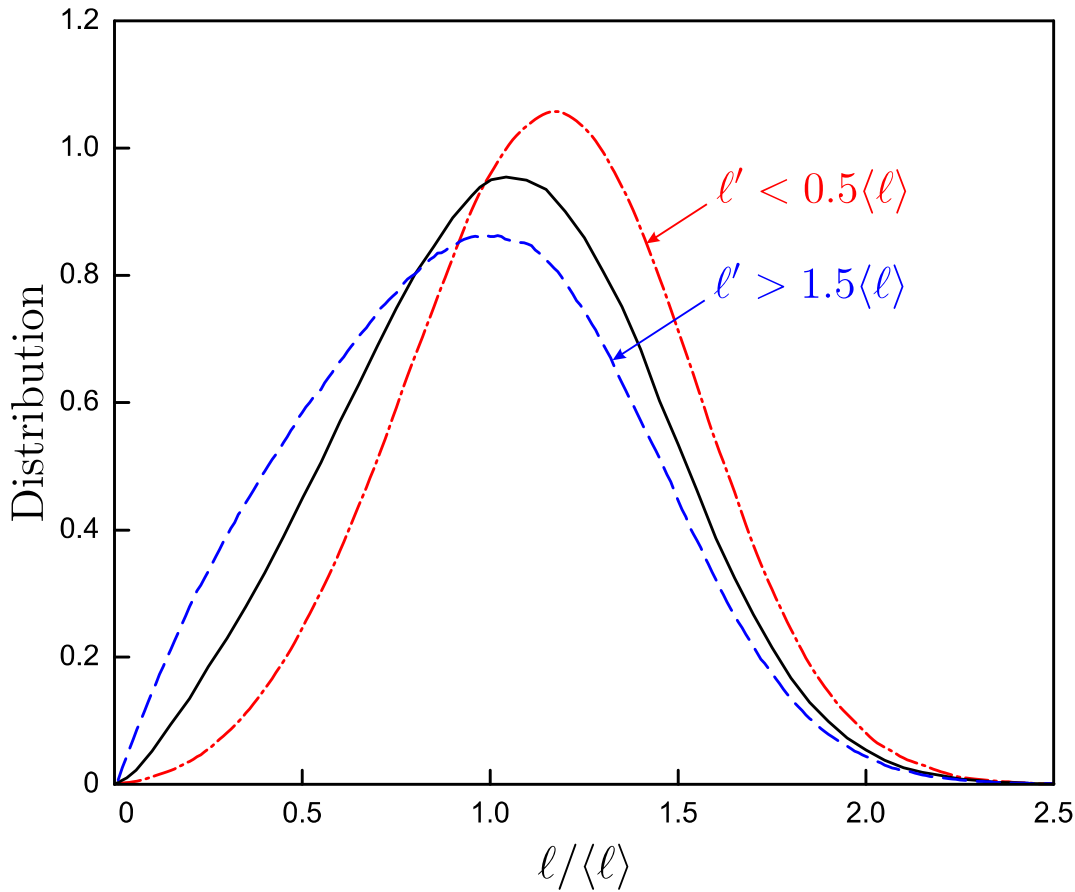}
\caption{Anti-correlation between adjacent gaps $\l$ and $\l'$ (for $\a=1$).  The unconstrained gap distribution of $\l$ (solid line) is compared to the gap distribution when the adjacent gap $\l'$ is unusually large, $\l'>1.5\av{\l}$, and unusually narrow, $\l'<0.5\av{\l}$.}
\label{correlations}
\end{figure}

Consider now correlations between adjacent gaps $\l$ and $\l'$.  We expect that the two gaps are strongly
anti-correlated: as the particle separating the two gaps moves one gap grows on expense of the other.  This anti-correlation is clearly visible in Fig.~\ref{correlations}, where we compare the (unconstrained) gap distribution to the gap distribution of $\l$ when $\l'<0.5\av{\l}$ and when $\l'>1.5\av{\l}$, for the case of $\a=1$.

Despite the intractable equation for $h_2$  it is possible to make a precise statement on the effect of correlations. 
We start with the {\em exact} equation
\begin{equation}
\label{n1n3}
\frac{\partial}{\partial t}n(\l,t)=-\frac{\partial}{\partial \l}\int_0^{\infty}\int_0^{\infty}
v(\l',\l,\l'')
n_3(\l',\l,\l'',t)\,d\l'd\l''\,,
\end{equation}
and use the scaling hypothesis (\ref{scaling}) to yield 
\begin{equation}
\label{h1}
2\a h(z)+\a z\frac{\partial}{\partial z}h(z)=\frac{\partial}{\partial z}\int_0^{\infty}\frac{dz_1}{z_1^\a} h_2(z,z_1)
-\frac{\partial}{\partial z}\frac{h(z)}{z^\a}\,.
\end{equation}
Integrating over $z$ we get
\begin{equation}
\a = \left[ h(z)\left\langle {1\over z_1^\alpha}\right\rangle_z - {h(z)\over z^\alpha} \right]^{z=\infty}_{z=0},
\label{eq_corr}
\end{equation}
where we have used the notation
\[
\left\langle {1\over z_1^\alpha}\right\rangle_z =
\int_0^\infty {dz_1\over z_1^\alpha} {h_2(z,z_1)\over h(z)}\,,
\]
for the conditional average of $z_1^{-\alpha}$ subject to the constraint that the interval $\l_1=z_1\bl$
is adjacent to an interval of length $\l=z\bl$.  It is also understood that  the limit of $z\to\infty$ needs changing to $z\to z_c$ if there is finite support.

Suppose that $\a>0$.  It is easy to realise that the first term does not contribute for $z=0$ and the second term
does not contribute for $z=\infty$. Therefore, we get
\begin{equation}
\a= \lim_{z\to\infty} h(z)\left\langle {1\over z_1^\alpha}\right\rangle_z 
+ \lim_{z\to 0} {h(z)\over z^\alpha} .
\label{exact}
\end{equation}
If one neglects correlations  the term $\av{z_1^{-\a}}_z$
does not depend on $z$, $h(z)$ vanishes for large $z$, and we found $h(z) \sim \a z^\alpha$ for small $z$, 
which agrees with Eq.~(\ref{small_z_MF}).

If we take correlations into account, we know from simulations that $h(z) \sim z^\beta$, with 
$\beta >\alpha$ (Fig.~\ref{fig.beta}), and the second term vanishes. 
Therefore, $\av{z_1^{-\a}}_z$ must diverge as $1/h(z)$ as $z\to\infty$.  Such behavior is believable, in view of
Fig.~\ref{correlations}, where we see that the conditional gap distribution when the adjacent gap is unusually large changes the $\beta$ exponent for the small-$z$ behavior.  The change in $\beta$ could result in divergence of the $(-\a)$-moment in the required way.

The following simple example serves to illustrate how such behavior may arise from anti-correlations.  Suppose that two adjacent gaps,
$x$ and $y$, are perfectly anti-correlated, such that their sum is fixed:
\[
h_2(x,y)\sim x(2-x)\delta(x+y-2)\,,\qquad 0\leq x,y\leq2\,,
\]
the $\delta$-function ensuring that $x+y=2$.  Integrating over $x$ we get $h(y)\sim y(2-y)$ (note that the example is symmetric; a similar $h(x)$ is obtained integrating over $y$).  Therefore, 
\[
\av{\frac{1}{x}}_y=\int_0^2 \frac{1}{x}\frac{h(x,y)}{h(y)}\,dx=\int_0^2 \frac{1}{x}\frac{x(2-x)}{y(2-y)}\delta(x+y-2)\,dx=\frac{1}{2-y}\,,
\]
which diverges in the same way as $1/h(y)$, as $y$ approaches the upper support limit of $y_c=2$.

Let us now analise Eq.~(\ref{eq_corr}) for negative $\a$. The terms evaluated at $z=\infty$
(or $z=z_c$) vanish and the term $h(z)z^{|\a|}$ vanishes at $z=0$ as well. Therefore, we have
\begin{equation}
h(0)\lim_{z\to 0}  \av{z_1^{|\a|}}_z =|\a| = h_{\hbox{\tiny MF}}(0)\av{z_1^{|\a|}}_{\hbox{\tiny MF}}\, ,
\end{equation}
where we have used Eq.~(\ref{small_z_MF}) for the last equality.  Thus, for $\a<0$, correlations revise the value of $h(0)$.

\comment{
Neglecting correlations, $\av{z_1^{-\a}}_z$ equals $\av{z_1^{-a}}$ and both sides of the equation are zero.
Because of correlations, however, the limits of the conditional average are non-trivial, and constrained as
implied.  For $\a\to-1$ we get $\av{z_1^{-\a}}_{z\to0}=\av{z_1^{-\a}}$, consistent with the idea that mean-field
becomes exact in that limit (see Fig~\ref{fig.beta}).
}

\subsection{Input}
\label{input.sect}
For homogeneous random input at rate $R$, we must add to the rhs of Eq.~(\ref{n1n3}) the terms
\[
R\left(-\l \,n(\l,t)+2\int_{\l}^{\infty}n(\l',t)\,d\l'\right)\,.
\]
The first term accounts for losses of a gap of length $\l$, which are proportional to $R\l$ and to the probability
for having the gap in the first place, while the second term accounts for the creation of gaps when a particle is deposited inside a gap $\l' >\l$, exactly at distance $\l$ from either edge.
Because of the randomizing effect of the input we expect that correlations play
a lesser role and the mean-field approximation does somewhat better in this case.

We specialize to the steady-state, setting the time derivative to zero, and use the mean-field ansatz
to obtain
\begin{equation}
\label{input.eq}
0=-\frac{1}{\a}\frac{\partial}{\partial z}\left[\left(\av{\frac{1}{z^\a}}-\frac{1}{z^\a}\right)h(z)\right]
-zh(z)+2\int_z^{\infty}h(z')\,dz'\,,
\end{equation}
where now $z=\l/\l_s$, ($\l_s\sim(A/R)^{1/(2+\a)}$).
The small-$z$ behaviour is very similar to that without input; $h(z)\sim\a\av{z}z^{\a}$ for $\a>0$,
and $h(z)\sim h(0)(1+z^{|\a|}/\av{z^{|\a|}})$, $h(0)=|\a|\av{z}/\av{z^{|\a|}}$, for $a<0$, while for large $z$
mean-field predicts $h(z)\sim\exp(-c_1z^2)$ for $\a>0$, and $h(z)\sim\exp(-c_2z^{2-|\a|})$ for $\a<0$
($c_1=\a$ and $c_2=|\a|/(2-|\a|)$ are constants).
These predictions seem to be borne out by simulations, although it is hard to make firm statements for the large-$z$ tail.  The large-$z$ tail of the distribution for $\a=1$ provides yet another difference from diffusion-limited
coalescence, where in the case of input $p(z)\sim \exp(-cz^{3/2})$~\cite{coal}.

\begin{figure}[ht]
\vspace*{0.cm}
\includegraphics*[width=0.99\textwidth]{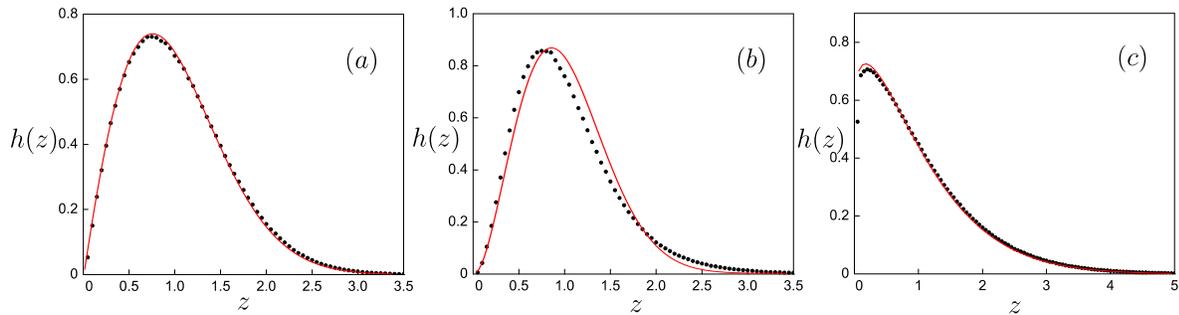}
\caption{Gap distributions for (a)~$\a=1$, (b)~$\a=2$, and (c)~$\a=-1/2$ in the case of input. Simulations (symbols) are compared to numerical integration of the mean-field equation~(\ref{input.eq}).  In all cases $z$
has been rescaled to yield $\av{z}=1$.}
\label{input.fig}
\end{figure}

In Fig.~\ref{input.fig} we compare simulations and numerical integration of the mean-field equation
for several values of $\a$.  The results are quite impressive, suggesting that input kills correlations very
effectively.

\section{Summary and Discussion}

In summary, we have analysed a class of deterministic reactions where nearest particles attract as $1/\l^\a$ and coalesce upon encounter.  The coarsening law is not
sufficient to distinguish between different reaction dynamics so we have focused instead on the distribution function $\rho(\l,t)$
for the gap $\l$ between nearest particles.  We have developed an exact Fokker-Planck hierarchy of equations
for the probabilities $\rho_m$ for finding $m$ adjacent gaps of given sizes, capitalizing on the fact that the $\rho_m$
provide a complete description of a one-dimensional particles system.  Truncating the hierarchy at the lowest level, we have studied the gap distribution at the mean-field approximation, neglecting all correlations between adjacent gaps.  While this provides a qualitative description of the gap distribution, important details, such as the $\beta$
exponent describing the small gap behavior $\rho(\l,t)\sim\l^{\beta}$, could not be captured in this way.  The role of correlations, however, was fully elucidated and we have found exact constraints that the distribution functions ought to obey due to their effect.  The mean-field approximation worked more satisfactorily for the case of random input, where the role of correlations is diminished, yielding the same $\beta$ exponent observed in simulations.

Perhaps the largest outstanding problem is how to proceed, analytically, beyond the lowest mean-field truncation
level, to obtain a more satisfying description of the gap distribution function and of the exponent $\beta$.  Especially intriguing is the numerical result that $\beta\approx1/2$ for $\a=0$: Could this be derived exactly, maybe by 
a totally different approach?  Cracking that question could suggest a more direct way of estimating $\beta$ for all
$\a$ values and perhaps for other deterministic reaction schemes.

\ack
We thank Maria Gracheva for generously allowing us access to her computer cluster and  
OG thanks Vladimir Privman for his constant guidance and support.
DbA thanks the NSF for partial funding of this project, and is also grateful to
the CNR for a Short-Term Mobility award from their International Exchange Program 
and for their warm hospitality while visiting Florence during the major part of this project.
PP acknowledges financial support from MIUR (PRIN 2007JHLPEZ).

\end{document}